\documentclass[12pt,twoside]{article}
\usepackage[english]{babel}
\usepackage{graphicx,rotating,epsfig}
\usepackage{a4p}
\usepackage{xspace}

\def\ra{\rightarrow} 
\def\GeV  {\ensuremath{\mathrm{ Ge\kern -0.1em V } }}
\def\GeVc2{\ensuremath{\mathrm{ Ge\kern -0.1em V }\kern -0.2em /c^2 }}
\def\MeVc2{\ensuremath{\mathrm{ Me\kern -0.1em V }\kern -0.2em /c^2 }}
\newcommand{\MT}{\ensuremath{M_{\mathrm{t}}}}
\newcommand{\MTll}{\ensuremath{\MT^{\mathrm{\ell\ell}}}}
\newcommand{\MTlj}{\ensuremath{\MT^{\mathrm{\ell\mbox{+}jets}}}}
\newcommand{\MTjj}{\ensuremath{\MT^{\mathrm{all-jets}}}}
\newcommand{\MTmet}{\ensuremath{\MT^{\mathrm{MEt}}}}
\newcommand{\MTcdf}{\ensuremath{\MT^{\mathrm{CDF}}}}
\newcommand{\MTdzero}{\ensuremath{\MT^{\mathrm{D0}}}}

\newcommand{\fb}{\ensuremath{\mathrm{fb}^{-1}}}
\newcommand{\ttbar}{\ensuremath{t\overline{t}}}
\newcommand{\WbWb}{\ensuremath{W^+ b W^- \overline{b}}}
\newcommand{\ljt}{\ensuremath{\ell\nu b q q^{\prime} \overline{b}}}
\newcommand{\had}{\ensuremath{q q^{\prime} b q q^{\prime} \overline{b}}}
\newcommand{\dil}{\ensuremath{\ell^{+}\nu b\ell^{-}\overline{\nu}\overline{b}}}
\newcommand{\ttljt}{\ensuremath{\ttbar\ra\WbWb\ra\ljt}}
\newcommand{\ttdil}{\ensuremath{\ttbar\ra\WbWb\ra\dil}}
\newcommand{\tthad}{\ensuremath{\ttbar\ra\WbWb\ra\had}}

\newcommand{\RunI}{\hbox{Run\,I}}
\newcommand{\RunII}{\hbox{Run\,II}}
\newcommand{\measStatSyst}[3]{\ensuremath{#1 \pm #2\thinspace(\textrm{stat}) \pm #3\thinspace(\textrm{syst})}\xspace}
\newcommand{\gevcc}[1]  {\ensuremath{#1~\mathrm{GeV}/c^{2}}}
\newcommand{\et}     {\ensuremath{E_{T}}\xspace}
\newcommand{\met}    {\mbox{$\protect \raisebox{.3ex}{$\not$}\et$}\xspace}

\newcommand{\pte}    {\ensuremath{p_{T}^{\mathrm{lep}}}}
\newcommand{\Lxy} {L_{XY}}
\begin{document}
%==============================================================================

\begin{center}
  {\LARGE FERMI NATIONAL ACCELERATOR LABORATORY}
\end{center}

\begin{flushright}
       FERMILAB-CONF-14-227-E-TD \\  % --- http://lss.fnal.gov/cgi-bin/getnumber.pl
       TEVEWWG/top2014/01 \\
       CDF Note 11105\\
          D\O\ Note 6444\\
       \vspace*{0.05in}
         July 2014 \\ 
\end{flushright}

\vskip 1cm

\begin{center}
  {\LARGE\bf 
    Combination of CDF and D\O\ results 
    on the mass of the top quark using up to \boldmath{$9.7\:\fb$} at
    the Tevatron\\
  }
  \vfill
  {\Large
    The Tevatron Electroweak Working Group\footnote{The Tevatron Electroweak 
    Working Group can be contacted at tev-ewwg@fnal.gov.\\  
    \hspace*{0.20in} More information can
    be found at {\tt http://tevewwg.fnal.gov}.} \\
    for the CDF and D\O\ Collaborations\\
  }
\end{center}
\vfill
\begin{abstract}
\noindent
  We summarize the current top-quark mass measurements from the CDF and
  D\O\ experiments at Fermilab.  We combine published
  \RunI\ (1992--1996) results with the most precise published and preliminary
  \RunII\ (2001--2011) measurements based on data corresponding to up to $9.7~\fb$ of $p\bar{p}$ collisions.
 Taking correlations of uncertainties into account, and
 combining the statistical and systematic uncertainties, the resulting
  preliminary Tevatron average mass of the top quark is $\MT =
  \gevcc{174.34 \ \pm \ 0.64}$, corresponding to a relative precision of $0.37\%$.
  
\end{abstract}

\vfill

%\setcounter{page}{2}

%\linenumbers

%==============================================================================
\section{Introduction}
\label{sec:intro}

This note reports the averaged mass of the top quark (\MT) obtained by
combining the most precise published and preliminary Tevatron measurements of
its mass at the Tevatron. This updates the combination presented in
Refs.~\cite{Mtop-tevewwgSum13} and ~\cite{TeVTopComboPRD}.  
Reference~\cite{TeVTopComboPRD} provides a detailed description  
of the systematic uncertainties.
In March 2014, the CDF, D\O\, ATLAS and CMS collaborations also provided a combination of their most precise
top-quark mass measurements~\cite{worldcombi}.

The CDF and D\O\ collaborations have performed several direct measurements of the
top-quark mass using data collected at the Tevatron
proton-antiproton collider located at the Fermi National Accelerator
Laboratory.  The pioneering measurements were first based on approximately
$0.1~\fb$ of \RunI\ data
~\cite{Mtop1-CDF-di-l-PRLa,Mtop1-CDF-di-l-PRLb,Mtop1-CDF-di-l-PRLb-E,Mtop1-D0-di-l-PRL,Mtop1-D0-di-l-PRD,Mtop1-CDF-l+jt-PRL,Mtop1-CDF-l+jt-PRD,Mtop1-D0-l+jt-old-PRL,Mtop1-D0-l+jt-old-PRD,Mtop1-D0-l+jt-new1,Mtop1-CDF-allh-PRL,Mtop1-D0-allh-PRL}
collected from 1992 to 1996
that included results from \tthad\ (all-jets),
\ttljt\ ($\ell$+jets), and 
\ttdil\ ($\ell\ell$) channels, where $\ell=e$ or $\mu$.  
Decays of $W \to \tau \nu_{\tau}$ followed by $\tau \to e, \mu$ are included in the direct $W \to e$ and
$W \to \mu$ channels. 
\RunII\ (2001--2011) had many top mass measurements, and
those considered in this paper are the most
recent results in these channels, using up to $9.3~\fb$ of data for CDF
(corresponding to the full CDF Run II data)
~\cite{
Mtop2-CDF-MEt-new,
Mtop2-CDF-l+jt-pub-new,
Mtop2-CDF-dil-2014,
Mtop2-CDF-allh-2014},
and up to $9.7~\fb$ of data for D\O\ (corresponding to the full D\O\ Run II data)
~\cite{ 
Mtop2-D0-l+jt-2014,
Mtop2-D0-di-l}.
The CDF analysis based upon charged-particle tracking for exploiting
the transverse decay length 
of $b$-tagged jets ($L_{XY}$) and the transverse momentum of
electrons and muons from $W$ boson decays ($\pte$) using
data corresponding to a luminosity of
1.9~$\fb$~\cite{Mtop2-CDF-trk},
has no plans for an update.

With respect to the March 2013 Tevatron combination \cite{Mtop-tevewwgSum13},  the \RunII\ CDF measurements in the 
$\ell\ell$ and all-jets
channels have been updated using their full \RunII\ data~\cite{Mtop2-CDF-dil-2014,Mtop2-CDF-allh-2014}.
The now published \RunII\ CDF measurements in the $\ell$+jets channel\,\cite{Mtop2-CDF-l+jt-pub-new} and $\met$+jets
(MEt) channel\,\cite{Mtop2-CDF-MEt-new} are unchanged. The measurement 
based on charged-particle tracking~\cite{Mtop2-CDF-trk} was
incorporated as described in the past 
combinations~\cite{Mtop-tevewwgSum11}. Possible correlation with the $\ell$+jets channel is discussed in that same 
reference.  
The \RunII\  D\O\ measurement in the 
$\ell$+jets
channel has been updated using the full \RunII\ data~\cite{Mtop2-D0-l+jt-2014}. 
This result is from Spring 2014 and was just accepted for publication in Phys. Rev. Lett.
The measurement in the 
$\ell\ell$ channel is  published and unchanged~\cite{Mtop2-D0-di-l}.

The Tevatron average \MT\ is obtained by combining five published
\RunI\ measurements~\cite{Mtop1-CDF-di-l-PRLb, Mtop1-CDF-di-l-PRLb-E,
  Mtop1-D0-di-l-PRD, Mtop1-CDF-l+jt-PRD, Mtop1-D0-l+jt-new1,
  Mtop1-CDF-allh-PRL} with five published \RunII\ 
results~\cite{Mtop2-CDF-MEt-new,Mtop2-CDF-l+jt-pub-new,
Mtop2-CDF-trk,Mtop2-D0-di-l,Mtop2-D0-l+jt-2014} and two preliminary \RunII\ CDF
results~\cite{Mtop2-CDF-dil-2014,Mtop2-CDF-allh-2014}.
This combination 
supersedes previous
combinations~\cite{Mtop-tevewwgSum13,Mtop-tevewwgSum11,Mtop1-tevewwg04,Mtop-tevewwgSum05,
  Mtop-tevewwgWin06,Mtop-tevewwgSum06, Mtop-tevewwgWin07, Mtop-tevewwgWin08, 
  Mtop-tevewwgSum08, Mtop-tevewwgWin09,Mtop-tevewwgSum10}. 

The definition and evaluation of the systematic uncertainties and of the
correlations among channels, experiments, and Tevatron runs is the outcome of many years of 
joint effort between the CDF and D\O\ collaborations that is described in detail
in Ref.~\cite{TeVTopComboPRD}.

In all measurements in the present combination, the analyses are calibrated to the Monte Carlo (MC)
top-quark mass definition. It is expected that the difference between the MC mass definition and the formal 
pole mass of the top quark is below 1~GeV~\cite{Buckley:2011ms}.

The input measurements and uncertainty categories used in the combination are 
detailed in Sections~\ref{sec:inputs} and~\ref{sec:uncertainty}, respectively. 
The correlations assumed in the combination are discussed in 
Section~\ref{sec:corltns}, and the resulting Tevatron average top-quark mass 
is given in Section~\ref{sec:results}.  A summary is presented
in Section~\ref{sec:summary}.
 
%==============================================================================
\section{Input Measurements}
\label{sec:inputs}

The twelve measurements of \MT\ used in this combination are shown in Table~\ref{tab:inputs}.
The \RunI\ measurements all have relatively large statistical
uncertainties and their systematic uncertainties are dominated by the
total jet energy scale (JES) uncertainty.  In \RunII\, both CDF and
D\O\ take advantage of the larger \ttbar\ samples available and employ
new analysis techniques to reduce both of these uncertainties.  In
particular, the \RunII\ D\O\ analysis in the $\ell$+jets channel and the 
\RunII\ CDF analyses in the $\ell$+jets, all-jets, and MEt channels 
constrain the response of light-quark jets using the kinematic information from $W\ra
qq^{\prime}$ decays (the so-called {\em in situ}
calibration)~\cite{Mtop1-CDF-l+jt-PRD,Abazov:2006bd}. 
Residual JES uncertainties associated with
$p_{T}$ and $\eta$ dependencies as well as uncertainties specific to
the response of $b$ jets are treated separately. The
\RunII\ D\O\ $\ell\ell$ measurement uses the JES determined in the
$\ell$+jets channel through the {\em in situ} calibration~\cite{Mtop2-D0-di-l}.

\vspace*{0.10in}

\begin{table}[t]
\caption[Input measurements]{Summary of the measurements used to determine the
  Tevatron average \MT.  Integrated luminosity ($\int \mathcal{L}\;dt$) has units of
  \fb, and all other numbers are in $\GeVc2$.  The uncertainty categories and 
  their correlations are described in Section~\ref{sec:uncertainty}.  The total systematic uncertainty 
  and the total uncertainty are obtained by adding the relevant contributions 
  in quadrature. The symbols ``n/a'' stands for ``not applicable'', ``n/e'' for ``not evaluated''.}
\label{tab:inputs}
\begin{center}

\renewcommand{\arraystretch}{1.30}
{\tiny
\begin{tabular}{l|ccc|cc|ccc|cc|cc} 
\hline \hline
       & \multicolumn{5}{c|}{{\RunI} published} 
       & \multicolumn{5}{c|}{{\RunII} published} 
       & \multicolumn{2}{c}{{\RunII} prel.}  \\ 
       & \multicolumn{3}{c|}{ CDF } 
       & \multicolumn{2}{c}{ D\O\ }
       & \multicolumn{3}{|c|}{ CDF }
       & \multicolumn{2}{c|}{ D\O\ }
       & \multicolumn{2}{c}{ CDF }
        \\

                      & $\ell$+jets & $\ell\ell$ &  all-jets & $\ell$+jets & $\ell\ell$ & $\ell$+jets &    Lxy  &    MEt    & $\ell$+jets & $\ell\ell$ &  $\ell\ell$ &  all-jets  \\
\hline                                                                                                                                                       
$\int \mathcal{L}\;dt$&       0.1   &     0.1    &    0.1   &     0.1     &     0.1    &      8.7    &    1.9  &     8.7   &    9.7      &        5.4 &       9.1   &      9.3   \\
\hline                                                                                                                                                       
Result                &   176.1     & 167.4      &186.0     & 180.1       & 168.4      &      172.85 &  166.90  &   173.93  &    174.98  &      174.00&    170.80   &    175.07 \\
\hline                                                                                                                                                       
\shortstack{ {\em In situ} light-jet cali- \\
bration (iJES)}       &        n/a  &      n/a   & n/a      &      n/a    &      n/a   &       0.49   &    n/a  &     1.05  &         0.41&       0.55 &       n/a  &      0.97  \\
\shortstack{ Response to $b$/$q$/$g$ \\
jets (aJES)}          &        n/a  &      n/a   & n/a      &      0.0    &      0.0   &       0.09   &    0.00 &     0.10  &         0.16&       0.40 &      0.18  &      0.02 \\

\shortstack{ Model for $b$ jets  \\
(bJES)}               &        0.6  &      0.8   & 0.6      &      0.7    &      0.7   &       0.16   &    0.00 &     0.17  &         0.09&       0.20 &      0.28  &      0.20  \\

 \shortstack{ Out-of-cone correction \\
     (cJES)}          &        2.7  &      2.6   & 3.0      &      2.0    &      2.0   &       0.21   &    0.36 &     0.18  &         n/a &        n/a &      1.65  &      0.37  \\
\shortstack{ Light-jet response (1) \\
 (rJES)}              &        3.4  &      2.7   & 4.0      &      n/a    &      n/a   &       0.48   &    0.24 &     0.40  &        n/a  &        n/a  &      1.72 &      0.42  \\
\shortstack{ Light-jet response (2) \\ 
    (dJES)}           &        0.7  &      0.6   & 0.3      &      2.5    &      1.1   &       0.07   &    0.06 &     0.04  &        0.21 &        0.56 &      0.46 &      0.09  \\
\shortstack{ Lepton modeling  \\
(LepPt)}              &        n/e  &      n/e   & n/e      &      n/e    &      n/e   &       0.03   &    0.00 &     n/a   &        0.01  &       0.35 &      0.36 &      n/a   \\
\shortstack{ Signal modeling \\
 (Signal)}            &        2.6  &      2.9   & 2.0      &      1.1    &      1.8   &       0.61   &    0.90 &     0.63  &        0.35  &       0.86 &      0.96 &     0.53\\
 \shortstack{ Jet modeling \\
  (DetMod)}           &        0.0  &      0.0   & 0.0      &      0.0    &      0.0   &       0.00   &    0.00 &     0.00  &        0.07  &       0.50 &      0.00  &      0.00 \\
 \shortstack{ $b$-tag modeling \\
  ($b$-tag)}           &        0.4  &      0.0   & 0.0      &      0.0    &      0.0   &       0.03   &    0.00 &     0.03  &        0.10  &       0.00 &      0.05  &      0.04 \\
\shortstack{ Background from \\
theory (BGMC)}        &        1.3  &      0.3   &     0.0  &      1.0    &      1.1   &       0.12   &    0.80 &     0.00  &        0.06  &       0.00 &      0.30  &      0.00   \\
 \shortstack{ Background based on\\
data (BGData)}        &        0.0  &      0.0   &    1.7   &      0.0    &      0.0   &      0.16    &    0.20 &     0.15  &        0.09  &       0.20 &      0.33  &      0.15  \\
 \shortstack{ Calibration method\\
(Method)}             &        0.0  &      0.7   & 0.6      &      0.6    &      1.1   &       0.05   &    2.50 &     0.21  &        0.07 &       0.51 &       0.19 &      0.87  \\
\shortstack{ Offset \hspace{0.5cm} \\
 (UN/MI)}             &        n/a  &      n/a   &    n/a   &      1.3    &      1.3   &       n/a    &      n/a&     n/a   &        n/a   &        n/a &      n/a   &      n/a  \\
 \shortstack{ Multiple interactions\\
 model (MHI)}         &        n/e  &      n/e   &    n/e   &      n/e    &      n/e   &       0.07   &    0.00 &     0.18  &         0.06&        0.00 &      0.30  &      0.22  \\
\hline                                                                                                                                                       
 \shortstack{ Systematic uncertainty\\
 (syst)}              &         5.3 &      4.9   &  5.7      &       3.9   &      3.6   &       0.98   &    2.82 &     1.36  &        0.63 &        1.49 &      2.69 &      1.55 \\
\shortstack{ Statistical uncertainty\\
 (stat)}               &        5.1 &     10.3   & 10.0      &       3.6   &     12.3   &       0.52   &    9.00 &     1.26  &        0.41 &        2.36 &      1.83 &      1.19 \\
\hline                                                                                                                                                       
Total uncertainty      &        7.3 &     11.4   &  11.5     &       5.3   &     12.8   &       1.12   &    9.43 &     1.85  &        0.76 &        2.80 &      3.26 &      1.95 \\ 
\hline
\hline
\end{tabular}
}
\end{center}
\end{table}

The D\O\ \RunII\ $\ell$+jets result is the  most accurate single result from the Tevatron and 
it is  accepted for publication~\cite{Mtop2-D0-l+jt-2014}.
The D\O\ \RunII\ $\ell\ell$ result is based on a neutrino weighting technique using 5.4~fb$^{-1}$
of data~\cite{Mtop2-D0-di-l}. 
Unlike the other inputs, the CDF all-jets measurement~\cite{Mtop2-CDF-allh-2014} uses a NLO generator as default program to
model \ttbar\ events (POWHEG~\cite{Frixione:2007nw}).

\vspace*{0.10in}

Table~\ref{tab:inputs} lists the individual uncertainties in each result,
subdivided into the categories described in the next Section.  The
correlations between the inputs are described in
Sec.~\ref{sec:corltns}.

%%%\clearpage

%==============================================================================
\section{Uncertainty Categories}
\label{sec:uncertainty}

We employ uncertainty categories similar to those used for the previous Tevatron
average~\cite{Mtop-tevewwgSum13,TeVTopComboPRD}, with small
modifications to better account for their correlations.
They are divided into sources of same or similar origin 
that are combined as in Ref.~\cite{TeVTopComboPRD}. 
For example, the {\it Signal modeling} ({\it Signal}) category
discussed below includes the uncertainties from  different systematic
sources that are correlated due to their origin in the modeling of the simulated signal.

Some systematic uncertainties have been separated into multiple
categories to accommodate specific types of correlations.
For example, the jet energy scale (JES) uncertainty is subdivided
into six components to more accurately accommodate our
best understanding of the relevant correlations among input measurements. 

In this note we use the new naming scheme described in Ref.~\cite{TeVTopComboPRD}. 
The previous names of the systematic uncertainties are also provided.  

%Each uncertainty category isdiscussed below.
\vspace*{0.10in}

\begin{description}
  \item[Statistical uncertainty (Statistics):] The statistical uncertainty associated with the
    \MT\ determination.
 \item[{\em In situ} light-jet calibration (iJES):] That part of the
   JES uncertainty that originates from
   {\em in situ} calibration procedures and is uncorrelated among the
   measurements.  In the combination reported here, it corresponds to
   the statistical uncertainty associated with the JES determination
   using the $W\ra qq^{\prime}$ invariant mass in the CDF \RunII\
   $\ell$+jets, all-jets, and  MEt measurements, and in the D\O\ Run~II
   $\ell\ell$ and $\ell$+jets
   measurements. 
   Residual JES uncertainties arising from effects
   not considered in the {\em in situ} calibration are included in other
   categories. 
   For the D\O\ Run~II $\ell\ell$ measurement, the uncertainty coming
   from transferring the $\ell$+jets calibration to the dilepton event topology
   is quoted in the {\it Light-jet response (2) (dJES)} category described below.
  \item[Response to \boldmath{$b/q/g$} jets (aJES):] That part of the JES
    uncertainty that originates from 
    average differences in detector electromagnetic over hadronic ($e/h$)
    response for hadrons produced in the fragmentation of $b$-quark and light-quark 
    jets. 
  \item[Model for \boldmath{$b$} jets (bJES):] That part of the JES uncertainty that originates from
    uncertainties specific to the modeling of $b$ jets and that is correlated
    across all measurements.  For both CDF and D\O\ this includes uncertainties 
    arising from 
    variations in the semileptonic branching fractions, $b$-fragmentation 
    modeling, and differences in the color flow between $b$-quark jets and light-quark
    jets.  These were determined from \RunII\ studies but back-propagated
    to the \RunI\ measurements, whose {\it Light-jet response (1)}  uncertainties ({\it rJES}, see below) were 
    then corrected to keep the total JES uncertainty constant.
  \item[Out-of-cone correction (cJES):] That part of the JES uncertainty that originates from
    modeling uncertainties correlated across all measurements.  
    It specifically includes the modeling uncertainties associated with light-quark 
    fragmentation and out-of-cone corrections. For D\O\ \RunII\ measurements,
    it is included in the {\it Light-jet response (2) (dJES)} category.
  \item[Light-jet response (1) (rJES):] The remaining part of the JES
    uncertainty that covers the absolute calibration for CDF's \RunI\
    and \RunII\ measurements. It also includes small contributions
    from the uncertainties associated with modeling multiple
    interactions within a single bunch crossing and corrections for
    the underlying event. 
  \item[Light-jet response (2) (dJES):] That part of the JES
    uncertainty that includes D\O's \RunI\ and \RunII\ calibrations of
    absolute response (energy dependent), the relative response
    ($\eta$-dependent), and the out-of-cone showering correction that 
    is a detector effect. This uncertainty term for CDF includes only
    the small relative response calibration ($\eta$-dependent) for
    \RunI\ and \RunII. 
  \item[Lepton modeling (LepPt):] The systematic uncertainty arising from uncertainties
    in the scale of lepton transverse momentum measurements. It was not
    considered as a source of systematic uncertainty in the \RunI\
    measurements. 
  \item[Signal modeling (Signal):] The systematic uncertainty arising from uncertainties
    in \ttbar\ modeling that is correlated across all
    measurements. This includes uncertainties from variations of the amount of initial and 
    final state radiation and from the choice of parton density function used
    to generate the \ttbar\ Monte Carlo samples
    that calibrate each method. When appropriate it also includes the uncertainty 
    from higher-order corrections evaluated from a comparison of \ttbar\ samples generated by {\textsc MC@NLO} ~\cite{MCNLO} and 
    {\textsc ALPGEN}~\cite{ALPGEN}, both interfaced to {\textsc HERWIG}~\cite{HERWIG5,HERWIG6} for the simulation of parton showers and hadronization.
    In this combination, the systematic uncertainty arising from a variation of the 
  phenomenological description of color reconnection (CR) between final state  particles \cite{CR,Skands:2009zm} 
  is included in the {\it Signal modeling} category.
  The CR uncertainty is obtained by taking the difference between the {\textsc PYTHIA}\,6.4 tune ``Apro" and the {\textsc PYTHIA}\,6.4 tune 
``ACRpro" that differ only in the CR model. 
In the latest analysis in the $\ell$+jets  channel, D\O\  uses  the  following samples: {\textsc PYTHIA} with Perugia 2011
versus Perugia 2011NOCR tunes  to estimate the uncertainty due to the  CR model. 
This uncertainty was not evaluated in Run~I since the Monte Carlo
generators available at that time did not allow for variations of the CR model.
These measurements therefore do not include this source of systematic uncertainty. Finally, the systematic 
uncertainty associated with variations of the MC generator used to calibrate the mass extraction method is added. 
It includes variations observed when 
    substituting {\textsc PYTHIA} %~[37--39]
    \cite{PYTHIA4,PYTHIA5,PYTHIA6} 
    (\RunI\ and \RunII) 
    or {\textsc ISAJET}~\cite{ISAJET} (\RunI) for {\textsc HERWIG}~\cite{HERWIG5,HERWIG6} when 
    modeling the \ttbar\ signal.  
  \item[Jet modeling (DetMod):] The systematic uncertainty arising from uncertainties 
in the modeling of jet interactions in the detector in the MC
simulation. For D\O\, this includes uncertainties from jet resolution
and identification. 
Applying jet algorithms to MC events, CDF finds that the resulting
efficiencies and resolutions closely match those in data. The small
differences propagated to $\MT$ lead to a negligible uncertainty of
0.005~GeV, which is then ignored.  

 \item[$b$-tag modeling ($b$-tag):] This is the part of the uncertainty related to the modelling of the $b$-tagging efficiency
 and the light-quark jet rejection factors in the MC simulation with respect to the data.

  \item[Background from theory (BGMC):] 
This systematic uncertainty on the background originating from theory
(MC) takes into account the    
 uncertainty in modeling the background sources. It is correlated between
    all measurements in the same channel, and includes uncertainties on the background composition, normalization, and shape of different components, e.g., the 
uncertainties from the modeling of the $W$+jets background in the $\ell$+jets channel  
associated with variations of the factorization scale used to simulate $W$+jets events. 

 \item[Background based on data (BGData):] This includes, among other sources,
    uncertainties associated with the modeling using data of the QCD
    multijet background in the all-jets,
    MEt, and $\ell$+jets channels and
    the Drell-Yan background in the $\ell\ell$ channel.
    This part is uncorrelated between experiments.

  \item[Calibration method (Method):] The systematic uncertainty arising from any source specific
    to a particular fit method, including the finite Monte Carlo statistics 
    available to calibrate each method. 
  \item[Offset (UN/MI):] This uncertainty is specific to D\O\ and includes the uncertainty
    arising from uranium noise in the D\O\ calorimeter and from the
    multiple interaction corrections to the JES.  For D\O\ \RunI\ these
    uncertainties were sizable, while for \RunII, owing to the shorter
    calorimeter electronics integration time and {\em in situ} JES calibration, these uncertainties
    are negligible.
  \item[Multiple interactions model (MHI):] The systematic uncertainty arising from a mismodeling of 
  the distribution of the number of collisions per Tevatron bunch crossing owing to the 
  steady increase in the collider instantaneous luminosity during data-taking. 
  This uncertainty has been separated from other sources to account for the fact that 
  it is uncorrelated between experiments.

\end{description}
These categories represent the current preliminary understanding of the
various sources of uncertainty and their correlations.  We expect these to 
evolve as we continue to probe each method's sensitivity to the various 
systematic sources with improving precision.

%==============================================================================
\section{Correlations}
\label{sec:corltns}

The following correlations are used for the combination:
\begin{itemize}
  \item The uncertainties in the {\it Statistical uncertainty (Stat)} and
    {\it Calibration method (Method)}
    categories are taken to be uncorrelated among the measurements.
  \item The uncertainties in the {\it In situ light-jet  
    calibration (iJES)}
    category are taken to be uncorrelated among the measurements
    except for D0's $\ell\ell$ and $\ell$+jets measurements, where
    this uncertainty is taken to be 100\% correlated since the
    $\ell\ell$ measurement uses the JES calibration  determined in
    $\ell$+jets channel.  
  \item The uncertainties in the {\it Response to $b$/$q$/$g$ jets (aJES)}, {\it Light-jet response (2) (dJES)}, {\it Lepton modeling (LepPt)}, {\it $b$-tag modeling ($b$-tag)}
    and {\it Multiple interactions model (MHI)} categories are taken
    to be 100\% correlated among all \RunI\ and all \RunII\ measurements 
    within the same experiment, but uncorrelated between \RunI\ and \RunII\
    and uncorrelated between the experiments.
  \item The uncertainties in the {\it Light-jet response (1) (rJES)}, {\it Jet modeling (DetMod)}, and {\it Offset (UN/MI)} categories are taken
    to be 100\% correlated among all measurements within the same experiment 
    but uncorrelated between the experiments.
  \item The uncertainties in the {\it Backgrounds estimated from theory (BGMC)} category are taken to be
    100\% correlated among all measurements in the same channel.
  \item The uncertainties in the {\it Backgrounds estimated from data (BGData)} category are taken to be
    100\% correlated among all measurements in the same channel and same run period, but uncorrelated between the experiments.
  \item The uncertainties in the {\it Model for $b$ jets (bJES)}, {\it Out-of-cone correction (cJES)}, and {\it Signal modeling (Signal)}
    categories are taken to be 100\% correlated among all measurements.
\end{itemize}
Using the inputs from Table~\ref{tab:inputs} and the correlations specified
here, the resulting matrix of total correlation coefficients is given in
Table~\ref{tab:coeff}.

\begin{table}[t]
\caption[Global correlations between input measurements]{The matrix of correlation coefficients used to determine the
  Tevatron average top-quark mass.}
\begin{center}
\renewcommand{\arraystretch}{1.30}
\tiny
\begin{tabular}{l|ccc|cc|ccc|cc|cc}
\hline \hline
       & \multicolumn{5}{c|}{{\RunI} published}
       & \multicolumn{5}{c|}{{\RunII} published}
       & \multicolumn{2}{c}{{\RunII} prel.}  \\
       & \multicolumn{3}{c|}{ CDF }
       & \multicolumn{2}{c}{ D\O\ }
       & \multicolumn{3}{|c|}{ CDF }
       & \multicolumn{2}{c|}{ D\O\ }
       & \multicolumn{2}{c}{ CDF }
        \\

                & $\ell$+jets & $\ell\ell$ & all-jets & $\ell$+jets & $\ell\ell$ & $\ell$+jets &  $\Lxy$   &   MEt   &  $\ell$+jets      & $\ell\ell$  &  $\ell\ell$   & all-jets  \\
\hline
CDF-I $\ell$+jets    &   1.00  &    0.29  &    0.32  &    0.26  &    0.11  &    0.49  &   0.07  &    0.26  &  0.19  &   0.12  &  0.54  &   0.27 \\
CDF-I $\ell\ell$     &   0.29  &    1.00  &    0.19  &    0.15  &    0.08  &    0.29  &   0.04  &    0.16  &  0.12  &   0.08  &  0.32  &   0.17 \\
CDF-I all-jets        &   0.32  &    0.19  &    1.00  &    0.14  &    0.07  &    0.30  &   0.04  &    0.16  &  0.08  &   0.06  &  0.37  &   0.18 \\
D\O-I $\ell$+jets    &   0.26  &    0.15  &    0.14  &    1.00  &    0.16  &    0.22  &   0.05  &    0.12  &  0.13  &   0.07  &  0.26  &   0.14 \\ 
D\O-I $\ell\ell$     &   0.11  &    0.08  &    0.07  &    0.16  &    1.00  &    0.11  &   0.02  &    0.07  &  0.07  &   0.05  &  0.13  &   0.07 \\
CDF-II $\ell$+jets   &   0.49  &    0.29  &    0.30  &    0.22  &    0.11  &    1.00  &   0.08  &    0.32  &  0.28  &   0.18  &  0.52  &   0.30 \\
CDF-II $\Lxy$        &   0.07  &    0.04  &    0.04  &    0.05  &    0.02  &    0.08  &   1.00  &    0.04  &  0.05  &   0.03  &  0.06  &   0.04 \\
CDF-II MEt           &   0.26  &    0.16  &    0.16  &    0.12  &    0.07  &    0.32  &   0.04  &    1.00  &  0.17  &   0.11  &  0.29  &   0.18 \\
D\O-II $\ell$+jets   &   0.19  &    0.12  &    0.08  &    0.13  &    0.07  &    0.28  &   0.05  &    0.17  &  1.00  &   0.36  &  0.15  &   0.14 \\
D\O-II $\ell\ell$    &   0.12  &    0.08  &    0.06  &    0.07  &    0.05  &    0.18  &   0.03  &    0.11  &  0.36  &   1.00  &  0.10  &   0.09 \\     
CDF-II $\ell\ell$    &   0.54  &    0.32  &    0.37  &    0.26  &    0.13  &    0.52  &   0.06  &    0.29  &  0.15  &   0.10  &  1.00  &   0.32 \\
CDF-II all-jets       &   0.27  &    0.17  &    0.18  &    0.14  &    0.07  &    0.30  &   0.04  &    0.18  &  0.14  &   0.09  &  0.32  &   1.00 \\

\hline
\hline
\end{tabular}
\end{center}
\label{tab:coeff}
\end{table}

The measurements are combined using a program implementing two 
independent methods: 
a numerical $\chi^2$ minimization and 
the analytic best linear unbiased estimator (BLUE) method~\cite{Lyons:1988, Valassi:2003}. 
The two methods are mathematically equivalent.
It has been checked that they give identical results for
the combination. The BLUE method yields the decomposition of the uncertainty on the Tevatron $\MT$ average in 
terms of the uncertainty categories specified for the input measurements~\cite{Valassi:2003}.

%==============================================================================
\section{Results}
\label{sec:results}

The resultant combined value for the top-quark mass is
\begin{eqnarray}
\nonumber
\MT=\gevcc{\measStatSyst{174.34}{0.37}{0.52}}. 
\end{eqnarray}
Adding the statistical and systematic uncertainties
in quadrature yields a total uncertainty of $\gevcc{0.64}$, corresponding to a
relative precision of 0.37\% on the top-quark mass.
The combination has a $\chi^2$ of 10.8 for 11 degrees of freedom, corresponding to
a probability of 46\%, indicating good agreement among all input
measurements.  The breakdown of the uncertainties is 
shown in Table\,\ref{tab:BLUEuncert}.  The total statistical and systematic  
uncertainties are reduced relative to the 
Spring  2013 combination \cite{Mtop-tevewwgSum13} and the published combination~\cite{TeVTopComboPRD} 
due to the new, most accurate,  D\O\ $\ell$+jets analysis and a small improvement 
due to  the CDF all-jets analysis. Both of these are based on the full data.

\begin{figure}[p]
\begin{center}
\includegraphics[width=0.8\textwidth]{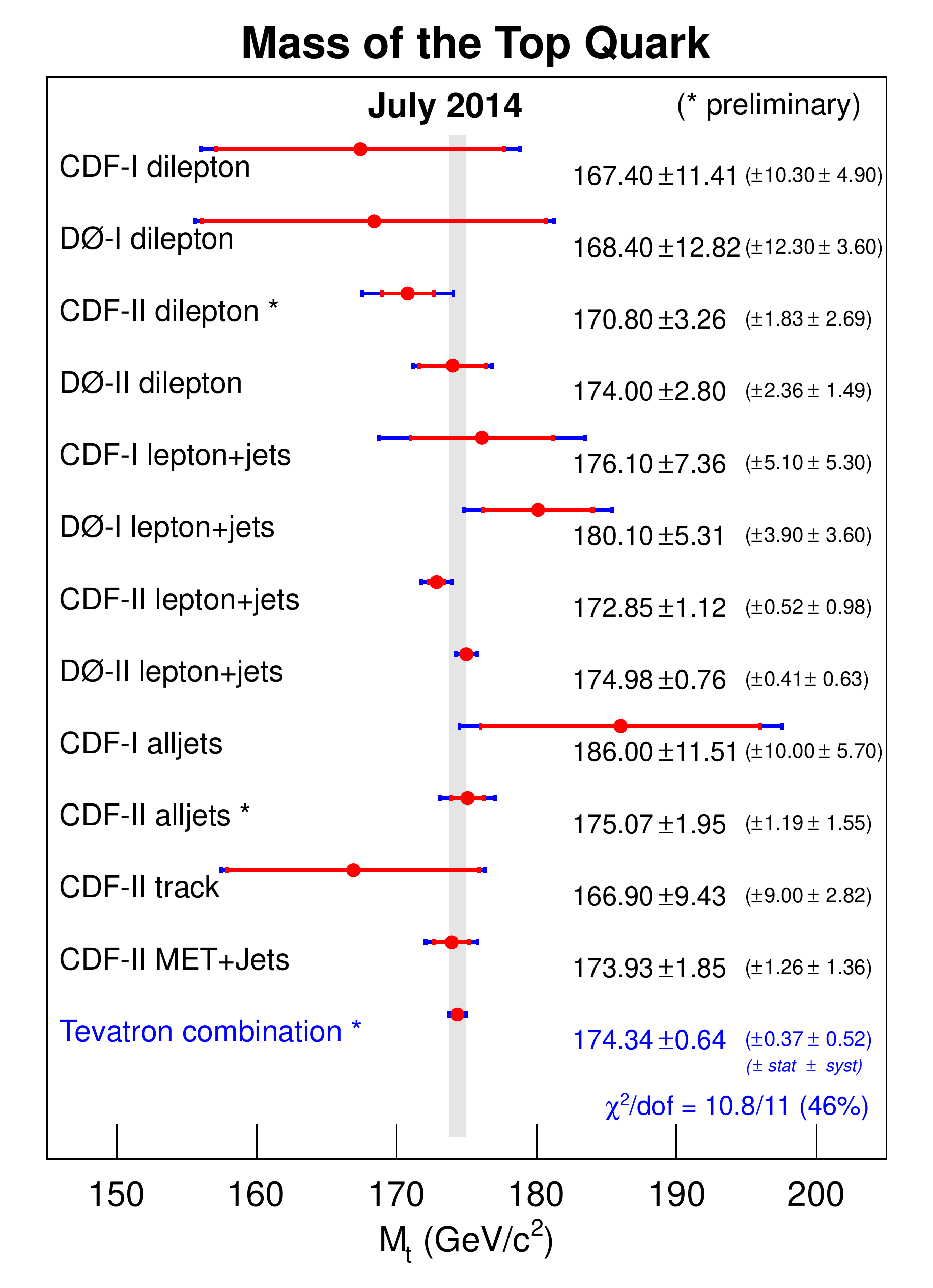}
\end{center}
\caption[Summary plot for the Tevatron average top-quark mass]
  {Summary of the input measurements and resulting Tevatron average
   mass of the top quark. The red lines correspond to the statistical uncertainty while the blue lines show
   the total uncertainty.}
\label{fig:summary} 
\end{figure}

The pull and weight for each of the inputs obtained from the
combination using the BLUE method, are listed in Table~\ref{tab:stat}.
The input measurements and the resulting Tevatron average mass of the top 
quark are summarized in Fig.~\ref{fig:summary}.
\vspace*{0.10in}

\begin{table}[tbh]
\caption{\label{tab:BLUEuncert} 
Summary of the Tevatron combined average $\MT$ and its uncertainties. The uncertainty categories are 
described in the text. The total systematic uncertainty and the total
uncertainty are obtained  
by adding the relevant contributions in quadrature.}
\begin{center}
\begin{tabular}{lc} \hline \hline
  & Tevatron combined values (GeV/$c^2$) \\ \hline
 $\MT$            & 174.34 \\ \hline

 {\em In situ} light-jet calibration (iJES)           &  0.31 \\
 Response to $b$/$q$/$g$ jets (aJES)                  &  0.10  \\
 Model for $b$ jets       (bJES)                  &  0.10 \\  
 Out-of-cone correction (cJES)                  &  0.02 \\ 
 Light-jet response (1) (rJES)                  &  0.05 \\
 Light-jet response (2) (dJES)                  &  0.13 \\
 Lepton modeling  (LepPt)                       &  0.07 \\
 Signal modeling  (Signal)                      &  0.34  \\
 Jet modeling (DetMod)                          &  0.03 \\
 $b$-tag modeling ($b$-tag)                         &  0.07 \\
 Background from theory (BGMC)                  &  0.04\\
 Background based on data (BGData)              &  0.08\\
 Calibration method (Method)                    &  0.07 \\
 Offset (UN/MI)                                 &  0.00 \\
 Multiple interactions model (MHI)              &  0.06 \\ \hline
 Systematic uncertainty (syst)                  &  0.52 \\ 
 Statistical uncertainty (stat)                 &  0.37 \\ \hline
 Total  uncertainty                             &  0.64 \\
   \hline \hline
\end{tabular}
\end{center}
\end{table}

\begin{figure}[p]
\begin{center}
\includegraphics[width=0.8\textwidth]{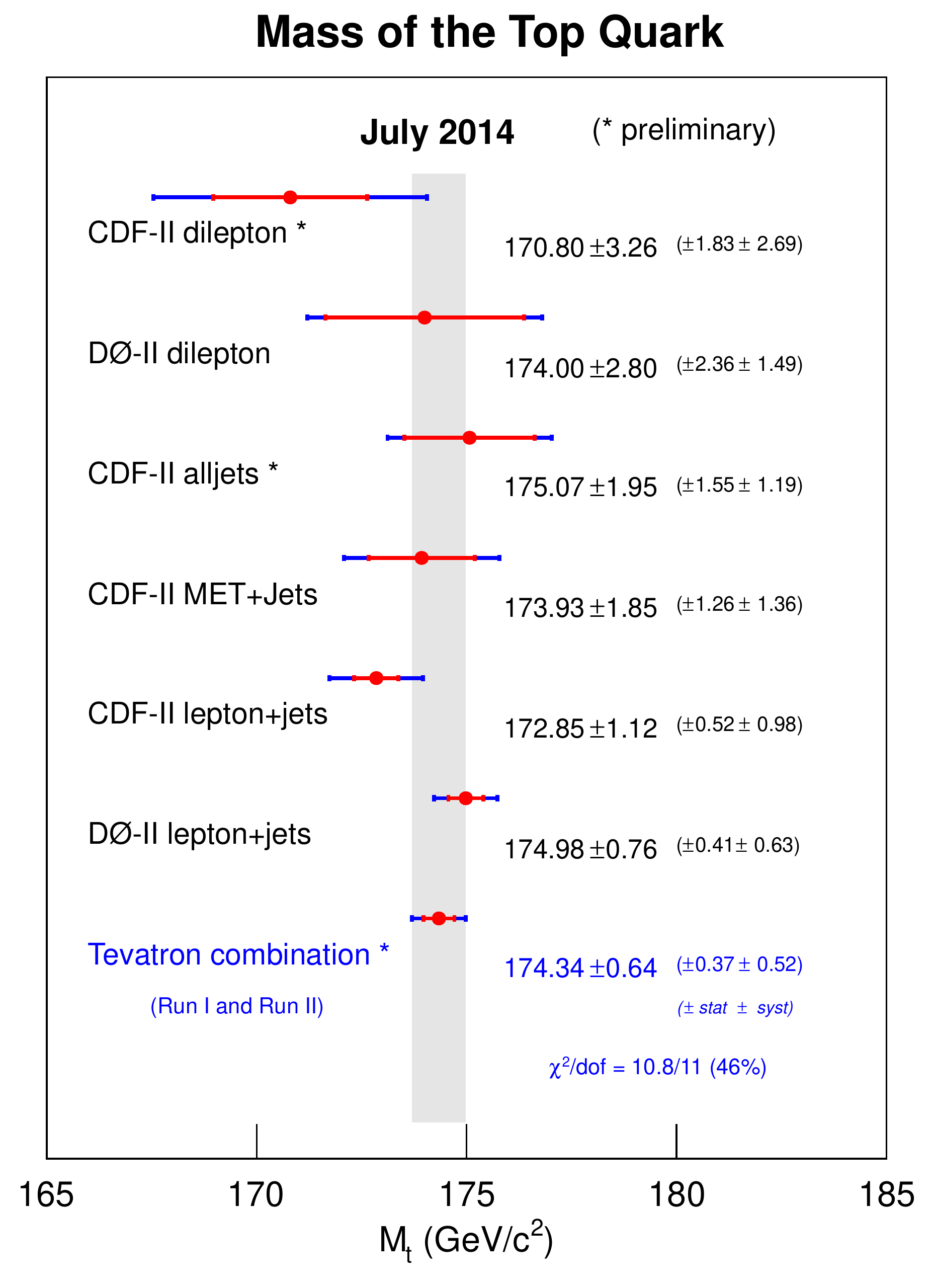}
\end{center}
\caption[Summary plot for the Tevatron average top-quark mass without RunI]
  {Summary of the input Run~II measurements and resulting Tevatron average
   mass of the top quark.}
\label{fig:summaryRunII} 
\end{figure}

\begin{figure}
\begin{center}
\includegraphics[width=0.9\textwidth]{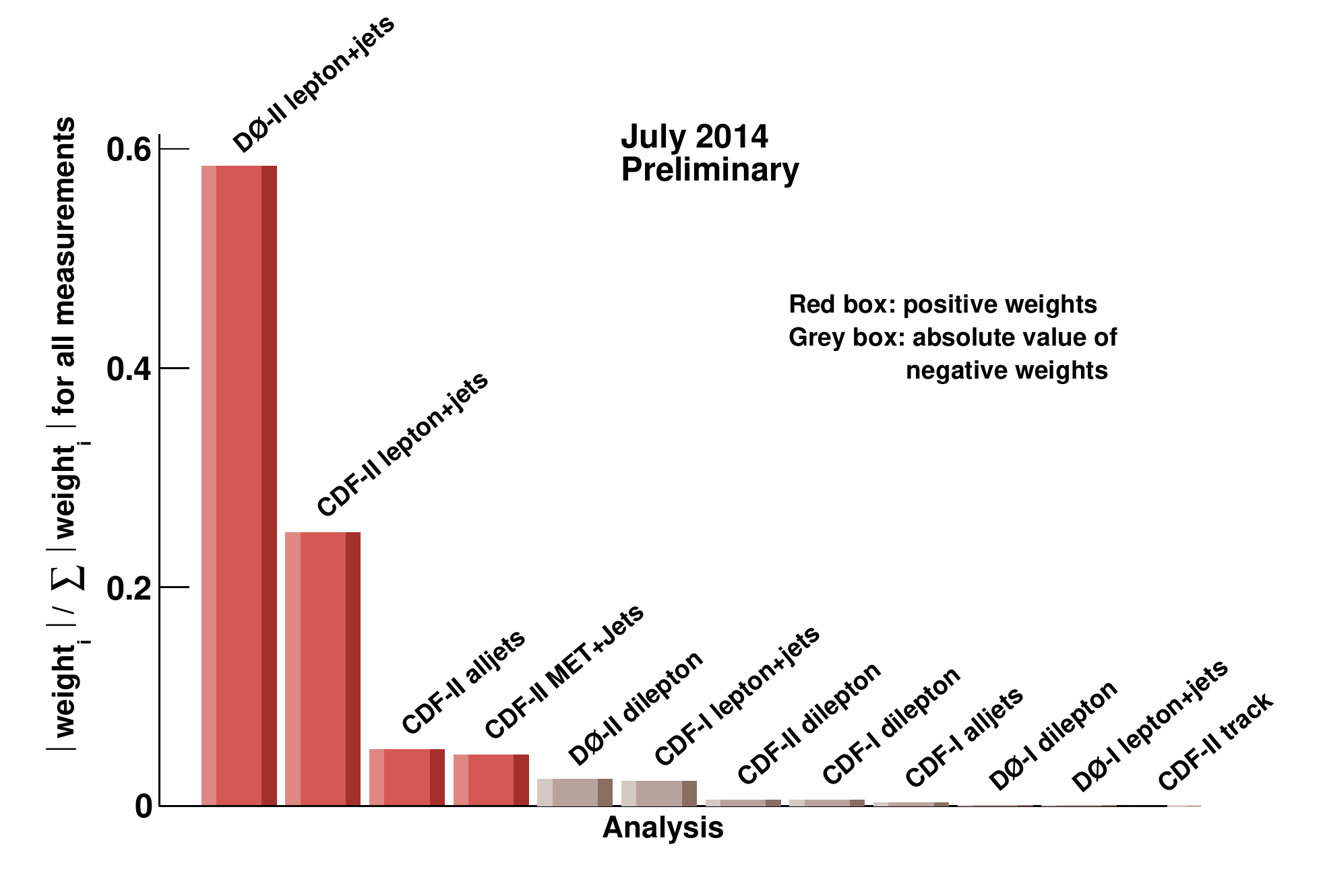}
\end{center}
\caption{Relative weights of the input measurements in the
  combination. The relative weights have been obtained by dividing the absolute value of each measurement weight by the sum over all measurements of the absolute values
of the weights. Negative weights are represented by their absolute
value, but using a grey color.}
\label{fig:Weights} 
\end{figure}

The weights of some of the measurements are negative, which occurs if
the correlation between two measurements 
is larger than the ratio of their total uncertainties.
In these instances the less precise measurement 
will acquire a negative weight.  While a weight of zero means that a
particular input is effectively ignored in the combination, channels
with a negative weight affect the resulting central value of $\MT$ and help reduce the total
uncertainty~\cite{Lyons:1988}. 
To visualize the weight that each measurement carries in the combination, Fig.\,\ref{fig:Weights} shows the
absolute values of the weight of each measurement divided by the sum of the absolute values of the weights
of all input measurements. Negative weights are represented by bins
with a different (grey) color. 
We note that due to correlations between the uncertainties, the relative weights
of the different input channels may be significantly different from
what would be expected from the total accuracy of each measurement represented by
error bars in Fig.~\ref{fig:summary}. A simular figure with only Run~II measurements, excluding CDF 
$L_{XY}$ measurement, is shown in  Fig.~\ref{fig:summaryRunII}

\begin{table}[t]
\caption[Pull and weight of each measurement]{The pull and weight for each of the
  inputs obtained from the combination with the BLUE method used to
  determine the average top-quark mass.}
\begin{center}
\renewcommand{\arraystretch}{1.30}
{\tiny
\begin{tabular}{l|ccc|cc|ccc|cc|cc}
\hline \hline
       & \multicolumn{5}{c|}{{\RunI} published}
       & \multicolumn{5}{c|}{{\RunII} published}
       & \multicolumn{2}{c}{{\RunII} prel.}  \\
       %\cline{2-12}
       & \multicolumn{3}{c|}{ CDF }
       & \multicolumn{2}{c}{ D\O\ }
       & \multicolumn{3}{|c|}{ CDF }
       & \multicolumn{2}{c|}{ D\O\ }
       & \multicolumn{2}{c}{ CDF }
        \\
       & $\ell$+jets & $\ell\ell$ & all-jets & $\ell$+jets & $\ell\ell$ & $\ell$+jets &  $\Lxy$   &   MEt   &  $\ell$+jets      & $\ell\ell$  &  $\ell\ell$   & all-jets  \\
\hline

Pull   & $0.24$  & $-0.61$  & $+1.01$  & $+1.09$  & $-0.46$  & $-1.64$  & $-0.791$  & $-0.24$  & $+1.60$   & $-0.13$ & $-1.11$ & $0.39$  \\
Weight [\%]  & $-2.6$ & $-0.7$ & $-0.4$ & $-0.1$ & $-0.14$ & $+28.8$ & $+0.1$ & $+5.5$ & $+67.2$ & $-2.9$ & $-0.66$ & $+6.0$ \\
\hline \hline
\end{tabular}
}
\end{center}
\label{tab:stat} 
\end{table} 

\begin{figure}[p]
\begin{center}
\includegraphics[width=0.8\textwidth]{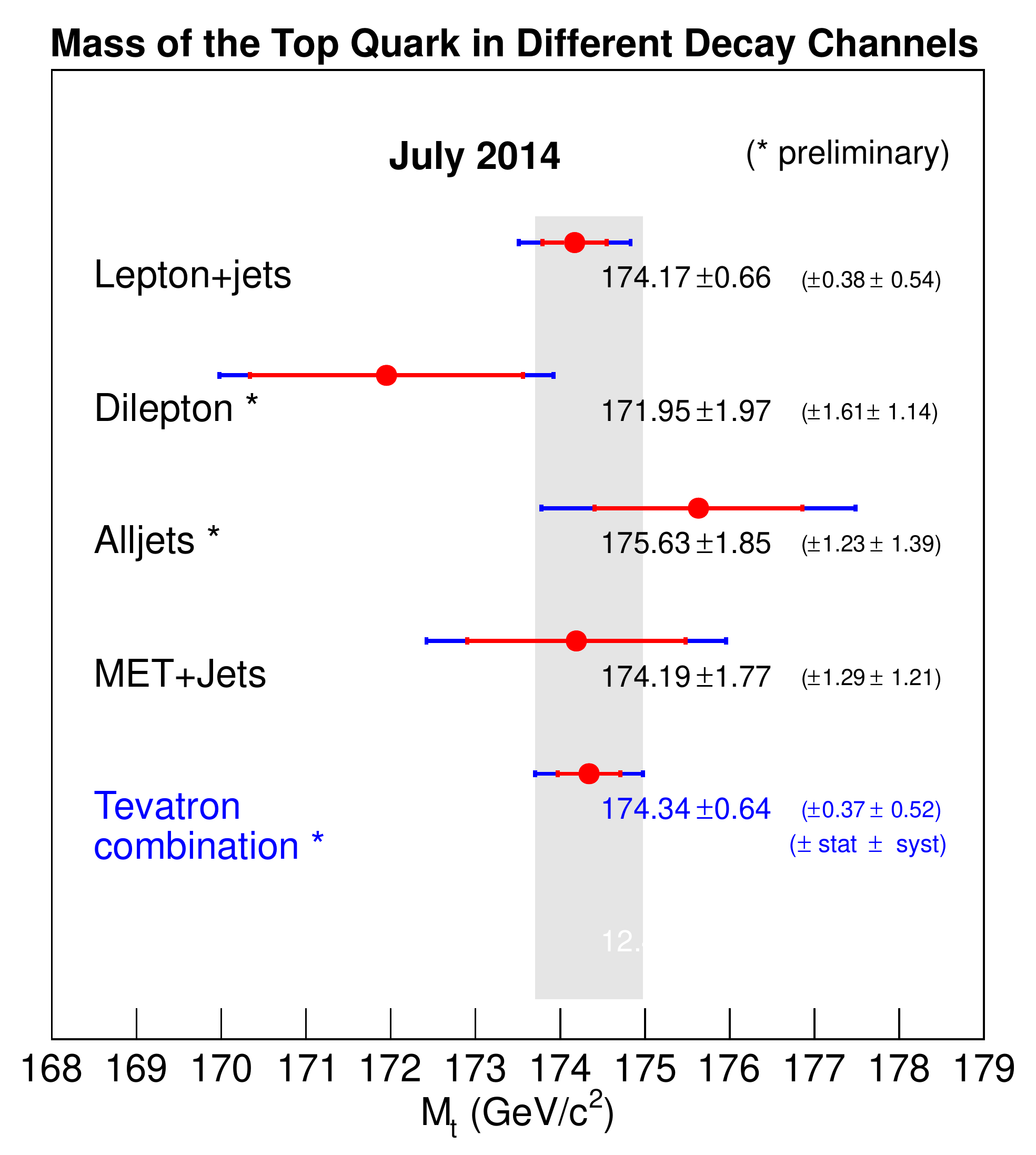}
\end{center}
\caption[Mtop in each channel]{Summary of the combination of the twelve
top-quark measurements by CDF and D\O\ for different final states.
The red lines correspond to the statistical uncertainty while the blue lines show
   the total uncertainty.}
\label{fig:three_observables} 
\end{figure}

No input has an anomalously large pull. 
Nevertheless, it is still
interesting to determine the mass separately 
in the all-jets, $\ell$+jets, $\ell\ell$, and MEt channels (leaving out
the $\Lxy$ measurement).
We use the same methodology,
inputs, uncertainty categories, and correlations as described above, but fit 
the four physical observables, \MTjj, \MTlj, \MTll, and \MTmet\ separately.
The results of these combinations are shown in
Fig.~\ref{fig:three_observables} and Table~\ref{tab:three_observables}.

Using the results of Table~\ref{tab:three_observables} 
 we calculate the following values including correlations:
$\chi^{2}(\ell+{\rm jets}-\ell\ell)=1.52/1$, $\chi^{2}(\ell+{\rm
  jets}-{\rm alljets})=0.64/1$,  $\chi^{2}(\ell+{\rm jets}-{\rm MEt})=0.00/1$, 
$\chi^{2}(\ell\ell-{\rm alljets})=2.34/1$,
  $\chi^{2}(\ell\ell-{\rm MEt})=0.87/1$, and $\chi^{2}({\rm
  alljets}-{\rm MEt})=0.36/1$.  
These correspond to probabilities 
of 22\%, 42\%, 99\%, 13\%, 35\%, and 55\%,  respectively, indicating
that the top-quark mass determined in each decay channel is consistent
in all cases.

\begin{table}[t]
\caption[Mtop in each channel]{Summary of the combination of the 12
measurements by CDF and D\O\ in terms of four physical quantities,
the mass of the top quark in the all-jets, $\ell$+jets,  $\ell\ell$, and MEt decay channels. }
\begin{center}
\renewcommand{\arraystretch}{1.30}
\begin{tabular}{ccrrrr}
\hline\hline

Parameter & Value (\GeVc2) & \multicolumn{4}{c}{Correlations} \\
               &                                 & $\MTjj$ &    $\MTlj$  &  $\MTll$ & $\MTmet$ \\ \hline
$\MTjj$ & $175.63 \pm 1.85$    & 1.00       &                   &      & \\
$\MTlj$ & $174.17 \pm 0.66$    & 0.21       &    1.00       &           &\\
$\MTll$ & $171.95 \pm 1.97$    & 0.21        &    0.41      & 1.00 &     \\
$\MTmet$& $174.19 \pm 1.77$    & 0.11        &    0.23      & 0.18 & 1.00 \\
\hline\hline
\end{tabular}
\end{center}
\label{tab:three_observables}
\end{table}

\begin{table}[!htb]
\caption[Mtop per experiment]{Summary of the combination of the 12
measurements by CDF and D\O\ in terms of two physical quantities,
the mass of the top quark measured in CDF and in D\O\ . }
\begin{center}
\renewcommand{\arraystretch}{1.30}
\begin{tabular}{ccrr}
\hline\hline

Parameter & Value (\GeVc2) & \multicolumn{2}{c}{Correlations} \\
               &                & $\MTcdf$ &    $\MTdzero$ \\ \hline
$\MTcdf$ & $173.12 \pm 0.92$    & 1.00       &          \\
$\MTdzero$  & $175.03 \pm 0.74$    & 0.25       &    1.00  \\
\hline\hline
\end{tabular}
\end{center}
\label{tab:two_observables}
\end{table}

In the same way, we can also fit two physical observables: the mass measured in CDF and the one
measured at D\O\ : \MTcdf\ and \MTdzero\ separately.
The result of these combinations is shown in Table~\ref{tab:two_observables}.
The chi-squared value including correlations is $\chi^{2}(CDF-D0)=3.5/1$ corresponding to a 
probability of 6.3\%.

%==============================================================================
\section{Summary}
\label{sec:summary}

An update was presented of the combination of measurements of the mass of the top quark
from the Tevatron experiments CDF and D\O.  This preliminary
combination includes five published \RunI\ measurements, seven published
\RunII\ measurements, and  
two preliminary \RunII\ measurements.  Most of these
measurements are  performed with the full data.  Taking into
account the statistical and systematic uncertainties and their
correlations, the preliminary result for the Tevatron average is
  $\MT=\gevcc{\measStatSyst{174.34}{0.37}{0.52}}$,
where the total uncertainty is obtained assuming Gaussian systematic uncertainties.
The central value is 1.14\,GeV/$c^2$  higher than the Tevatron Spring 2013
average~\cite{Mtop-tevewwgSum13} of 
$\MT=173.20\pm0.87$\,GeV/$c^2$.
Adding in quadrature the statistical and systematic uncertainties
yields a total uncertainty of $\gevcc{0.64}$ that represents an
improvement of $26\%$.
Compared to the world average, the central value of this combination is $\gevcc{1.00}$ 
higher and its precision is 16\% better.  

The mass of the top quark is now known with a relative precision of
0.37\%, limited by the systematic uncertainties, which are dominated by
the uncertainty on {\em in situ} calibration and signal modeling.  
This result will be further improved when all analysis channels from  
CDF and D\O\ using the full Run~II data 
are optimized in their treatment of their systematic uncertainties.

\section{Acknowledgments}
\label{sec:ack}

We thank the Fermilab staff and the technical staffs of the
participating institutions for their vital contributions. 
This work was supported by  
DOE and NSF (USA),
CONICET and UBACyT (Argentina), 
CNPq, FAPERJ, FAPESP and FUNDUNESP (Brazil),
CRC Program, CFI, NSERC and WestGrid Project (Canada),
CAS and CNSF (China),
Colciencias (Colombia),
MSMT and GACR (Czech Republic),
Academy of Finland (Finland),
CEA and CNRS/IN2P3 (France),
BMBF and DFG (Germany),
Ministry of Education, Culture, Sports, Science and Technology (Japan), 
World Class University Program, National Research Foundation (Korea),
KRF and KOSEF (Korea),
DAE and DST (India),
SFI (Ireland),
INFN (Italy),
CONACyT (Mexico),
NSC(Republic of China),
FASI, Rosatom and RFBR (Russia),
Slovak R\&D Agency (Slovakia), 
Ministerio de Ciencia e Innovaci\'{o}n, and Programa Consolider-Ingenio 2010 (Spain),
The Swedish Research Council (Sweden),
Swiss National Science Foundation (Switzerland), 
FOM (The Netherlands),
STFC and the Royal Society (UK),
and the A.P. Sloan Foundation (USA).

\clearpage

\providecommand{\href}[2]{#2}\begingroup\raggedright\endgroup

\end{document}